\documentclass[onecolumn, trackchanges]{aastex631}
\usepackage[utf8]{inputenc}
\usepackage[T1]{fontenc}
\usepackage[normalem]{ulem}

\received{Apr 20, 2023}
\revised{May 18, 2023}
\accepted{May 20, 2023}
\submitjournal{ApJ}

\shorttitle{Quantifying Acceleration}
\shortauthors{Halekas et al.}

\graphicspath{{./}{figures/}}

\begin{document}

\title{Quantifying the Energy Budget in the Solar Wind from 13.3-100 Solar Radii}

\correspondingauthor{Jasper S. Halekas}
\email{jasper-halekas@uiowa.edu}

\author[0000-0001-5258-6128]{J.~S. Halekas}
\affil{Department of Physics and Astronomy, 
University of Iowa, 
Iowa City, IA 52242, USA}

\author[0000-0002-1989-3596]{S.~D. Bale}
\affil{Space Sciences Laboratory, University of California, Berkeley, CA 94720, USA}
\affil{Physics Department, University of California, Berkeley, CA 94720, USA}

\author[0000-0001-6235-5382]{M. Berthomier}
\affil{Laboratoire de Physique des Plasmas, CNRS, Sorbonne Universite, Ecole Polytechnique, Observatoire de Paris, Universite Paris-Saclay, Paris, 75005, France}

\author[0000-0003-4177-3328]{B.~D.~G. Chandran}
\affil{Department of Physics \& Astronomy, University of New Hampshire, Durham, NH 03824, USA}
\affil{Space Science Center, University of New Hampshire, Durham, NH 03824, USA}

\author[0000-0002-9150-1841]{J.~F. Drake}
\affil{Department of Physics, University of Maryland, College Park, MD 20742, USA}
\affil{Institute for Physical Science and Technology, University of Maryland, College Park, MD 20742, USA}
\affil{Joint Space Science Institute, University of Maryland, College Park, MD 20742, USA}

\author[0000-0002-7077-930X]{J.~C. Kasper}
\affiliation{BWX Technologies, Inc., Washington DC 20002, USA}
\affiliation{Climate and Space Sciences and Engineering, University of Michigan, Ann Arbor, MI 48109, USA}

\author[0000-0001-6038-1923]{K.~G. Klein}
\affiliation{Lunar and Planetary Laboratory, University of Arizona, Tucson, AZ 85719, USA.}

\author[0000-0001-5030-6030]{D.~E. Larson}
\affil{Space Sciences Laboratory, University of California, Berkeley, CA 94720, USA}

\author[0000-0002-0396-0547]{R. Livi}
\affil{Space Sciences Laboratory, University of California, Berkeley, CA 94720, USA}

\author[0000-0002-1573-7457]{M.~P. Pulupa}
\affil{Space Sciences Laboratory, University of California, Berkeley, CA 94720, USA}

\author[0000-0002-7728-0085]{M.~L. Stevens}
\affil{Smithsonian Astrophysical Observatory, Cambridge, MA 02138, USA}

\author[0000-0003-1138-652X]{J.~L. Verniero}
\affil{NASA/Goddard Space Flight Center, Greenbelt, MD 20771, USA}

\author[0000-0002-7287-5098]{P. Whittlesey}
\affil{Space Sciences Laboratory, University of California, Berkeley, CA 94720, USA}

\begin{abstract}
A variety of energy sources, ranging from dynamic processes like magnetic reconnection and waves to quasi-steady terms like the plasma pressure, may contribute to the acceleration of the solar wind. We utilize a combination of charged particle and magnetic field observations from the Parker Solar Probe (PSP) to attempt to quantify the steady-state contribution of the proton pressure, the electric potential, and the wave energy to the solar wind proton acceleration observed by PSP between 13.3 and $\sim \! 100$ solar radii ($R_{\sun}$). The proton pressure provides a natural kinematic driver of the outflow. The ambipolar electric potential acts to couple the electron pressure to the protons, providing another definite proton acceleration term. Fluctuations and waves, while inherently dynamic,  can act as an additional effective steady-state pressure term. To analyze the contributions of these terms, we utilize radial binning of single-point PSP measurements, as well as repeated crossings of the same stream at different distances on individual PSP orbits (i.e. “fast radial scans”). In agreement with previous work, we find that the electric potential contains sufficient energy to fully explain the acceleration of the slower wind streams. On the other hand, we find that the wave pressure plays an increasingly important role in the faster wind streams. The combination of these terms can explain the continuing acceleration of both slow and fast wind streams beyond 13.3 $R_{\sun}$.   
\end{abstract}

\keywords{}

\section{Introduction} \label{sec:intro}

Since the prediction \citep{parker_dynamics_1958} and rapid subsequent discovery \citep{gringauz_study_1960, neugebauer_solar_1962} of a supersonic plasma wind flowing out from the Sun over sixty years ago, considerable theoretical and experimental effort has gone towards explaining the acceleration of the solar wind. The earliest theoretical work by Parker \citep{parker_dynamics_1958,parker_dynamical_1965} focused on the acceleration by the plasma pressure gradient, under various simplifying assumptions such as isothermality and purely radial and unmagnetized plasma flow. Extensions of the basic Parker model have incorporated additional physics, such as polytropic models for the plasma \citep{parker_hydrodynamic_1960, shi_acceleration_2022, dakeyo_statistical_2022}, non-radial magnetic fields \citep{weber_angular_1967, urch_model_1969}, and the effects of field line expansion between the corona and the sonic critical point \citep{holzer_conductive_1980, wang_solar_1990}. Given that even the earliest observations of the solar wind clearly revealed discrete streams with a wide range of wind speeds flowing from different sources on the Sun \citep{neugebauer_solar_1962}, a fully successful solar wind model (or models) should explain both the slow and fast wind. Unfortunately, while pressure-driven fluid wind models have had great success in explaining the slow solar wind, they have difficulty explaining the fast wind, at least for realistic coronal temperatures \citep{leer_acceleration_1982}. 

Since the solar wind is only weakly collisional during most of its expansion, significant electric fields must develop in order to maintain quasi-neutrality in the flowing plasma \citep{pannekoek_ionization_1922}. To maintain quasi-neutrality globally, the electric potential must not only balance the electron and ion density, but also balance the electron and ion currents \citep{jockers_solar_1970,lemaire_kinetic_1971}. Given the similar electron and ion temperatures, but highly unequal masses, this requires an electric potential several times larger than the electron thermal energy, on the order of several hundred volts. Recognizing the importance of this electric potential, researchers have developed a class of collision-less “exospheric” models that rely entirely on the electric field to accelerate the ions \citep{lemaire_kinetic_1973}. The exospheric models can successfully explain the slow solar wind. However, like the pressure-driven models, they have difficulty in explaining the fast wind, unless a suprathermal coronal electron population \citep{scudder_causes_1992} sufficient to generate the necessary electric field exists \citep{maksimovic_kinetic_1997, pierrard_lorentzian_1996, zouganelis_transonic_2004}. 

We can recognize a close relationship between the exospheric models and the pressure-driven fluid models, given that the parallel electric field in the solar wind effectively couples the electron pressure gradient to the ions. We can identify this electric field as the ambipolar term in a generalized Ohm’s law, and in the appropriate limit one can show an equivalence between the electron pressure gradient force in the fluid models and the electric field force in the exospheric models \citep{parker_kinetic_2010}. 

Given these considerations, the fast wind seemingly requires an additional energy source, above and beyond the coronal plasma pressure and the associated electric potential \citep{leer_acceleration_1982, hansteen_solar_2012}. The convective and wave energy in the solar photosphere and chromosphere provide obvious (and enormous) sources of energy \citep{de_pontieu_chromospheric_2007, cranmer_role_2015, cranmer_self-consistent_2012, mccomas_understanding_2007, zank_turbulence_2021}. Many fast wind models ultimately rely on tapping this energy. One plausible class of fast wind models relies on magnetic reconnection, presumably in the form of interchange reconnection between open and closed magnetic field lines \citep{fisk_acceleration_1999, fisk_acceleration_2003, owens_signatures_2020, bale_interchange_2022, raouafi_magnetic_2023}. Another class of fast wind models relies on the energy carried in waves and/or turbulent fluctuations  \citep{parker_dynamical_1965, belcher_alfvenic_1971, jacques_momentum_1977, zank_waves_1992, matthaeus_evolution_1994, tu_two-fluid_1997, matthaeus_coronal_1999, cranmer_self-consistent_2007, holst_data-driven_2010, verdini_turbulence-driven_2009, chandran_incorporating_2011, zank_theory_2017, zank_theory_2018, zank_spectral_2020,  reville_role_2020, adhikari_solar_2020, adhikari_evolution_2021, chandran_approximate_2021, perez_how_2021, zank_turbulence_2021, telloni_first_2023}, with these waves effectively providing an additional pressure term that, if coupled to the plasma, provides additional acceleration. Solar wind models powered by reflection of outward-propagating Alfv\'{e}n waves and/or by energy from the constantly upwelling magnetic carpet of the lower corona show great promise for providing a fully self-consistent theory of solar wind acceleration \citep{zank_turbulence_2021}. The nearly incompressible turbulence solar wind models represent one successful class of wave/turbulence driven models \citep{zank_theory_2017, zank_theory_2018, zank_spectral_2020}. This type of turbulence-driven model has had significant recent success in reproducing solar wind observations from PSP and Solar Orbiter in both the slow and fast solar wind \citep{adhikari_solar_2020, adhikari_evolution_2021, telloni_first_2023}

Observations from a multitude of spacecraft have provided observational constraints on the nature of the solar wind. Even the first early Mariner 2 observations  revealed the dichotomous nature of the solar wind, with slow streams interspersed with fast streams \citep{neugebauer_solar_1962}. The polar-orbiting Ulysses mission revealed both the latitudinal structure and the solar cycle variation of the solar wind. Ulysses observations suggested that the fast wind emanates from the cooler coronal holes, while slow wind primarily comes from the warmer equatorial “streamer-belt” regions of the Sun \citep{phillips_ulysses_1995, mccomas_ulysses_1998, mccomas_solar_2000, mccomas_ulysses_2001, abbo_slow_2016}. The resulting apparent inverse relationship between coronal electron temperature and bulk speed, confirmed by ion charge state measurements \citep{geiss_southern_1995}, seemingly contradicts the predictions of both fluid and exospheric models, but may prove consistent with models involving magnetic reconnection \citep{fisk_acceleration_2003} or Alfv\'{e}n wave driven models \citep{chandran_approximate_2021}. Helios 1 and 2 gave us our first view of the wind closer to the Sun, revealing the pervasive non-thermal nature of both the electron and ion distributions near the Sun (which persists to some degree even out to 1 AU) \citep{rosenbauer_survey_1977, pilipp_characteristics_1987, phillips_anisotropic_1989, pilipp_large-scale_1990, marsch_kinetic_2006}, and the higher level of Alfv\'{e}nic fluctuations in the plasma \citep{roberts_origin_1987}. In situ observations from various spacecraft near the Earth, notably including Wind and ACE, have provided further important constraints on the kinetic microstate and composition of the plasma \citep{bale_magnetic_2009, kasper_windswe_2002, lepri_solar_2013, wilson_iii_quarter_2021, salem_precision_2021}, from which we have inferred many details about the processes that have acted on the wind during its expansion from the Sun.

The Parker Solar Probe (PSP) \citep{fox_solar_2016} is now making in situ measurements even closer to the Sun than the Helios perihelion. The more pristine nature of the charged particle distributions closer to the Sun has enabled a clearer view of the electron temperature-bulk speed anticorrelation \citep{maksimovic_anticorrelation_2020, halekas_electrons_2020}. PSP has also measured pervasive and striking non-thermal features in the charged particle distributions. Among other things, PSP observations have revealed a truncation in the sunward portion of the suprathermal electron velocity distribution function \citep{bercic_ambipolar_2021, halekas_sunward_2021}, which we interpret as representing the boundary in phase space between electrostatically trapped electrons and escaping electrons; the energy of this boundary thus corresponds to the local value of the electric potential with respect to infinity. Recent work suggests that this electric potential, extrapolated to the corona, contains enough energy to provide the entire acceleration of the slowest wind streams \citep{halekas_radial_2022}. On the other hand, the electric potential can only provide a small fraction of the total acceleration of the faster streams. Meanwhile, PSP and Solar Orbiter observations have dramatically shown the widespread presence of highly coherent large-amplitude Alfv\'{e}nic fluctuations, often termed “switchbacks”, near the Sun \citep{bale_highly_2019, kasper_alfvenic_2019, telloni_observation_2022, raouafi_parker_2023}. These structures occur in both the fast and slow wind near the Sun, but the magnitude of the velocity fluctuation scales with the Alfv\'{e}n speed \citep{matteini_ion_2015}, and their amplitude is thus typically larger in faster wind. The energy contained in these fluctuations  also provides an important energy source for the acceleration of the wind, given that processes exist to couple it to the bulk flow. 

One potential method of revealing the energy sources of solar wind acceleration involves tracking the contribution of different forms of solar wind energy flux along an expanding flux tube as the wind flows outward and accelerates. This method has been fruitfully utilized to analyze both simulations and observations \citep{leer_energy_1980, schwartz_radial_1983, marsch_helios_1984,   chandran_incorporating_2011, chen_evolution_2020}. While quantifying the energy flux terms does not reveal the actual mechanisms that convert one form of energy to another, it can at least demonstrate that such conversion occurs. In this work, we attempt to quantify the energy budget available for solar wind acceleration by utilizing radial binning of single-point PSP measurements from orbits 4-13, as well as repeated crossings of the same stream at different distances on individual PSP orbits.

\section{Radial Evolution of Energy Flux Terms} \label{sec:radial}

To characterize the steady-state acceleration of the solar wind, we analyze the parallel proton energy flux. Assuming expansion purely along the magnetic field, we write the kinetic energy and enthalpy flux of the protons as
\begin{equation}
    F_{proton}=n U \left( \frac{m_p U^2}{2}+k_B T_{\perp p} + \frac{3 k_B T_{\parallel p}}{2} \right).
\end{equation}
Here $m_p$ and $n$ are the proton mass and number density, $U$ is the bulk speed, and $T_{\perp}$ and $T_{\parallel}$ are the temperature components perpendicular and parallel to the magnetic field. To compute the proton energy flux terms, we utilize a combination of data from the Solar Wind Electrons, Alphas, and Protons (SWEAP) experiment \citep{kasper_solar_2016} and the FIELDS experiment \citep{bale_fields_2016} on PSP. Near perihelion on orbits 4-13 (and later orbits not included), the Solar Probe Analyzer-Ions (SPAN-I) sensor \citep{livi_solar_2022} measures the majority of the proton velocity distribution, and we therefore utilize the SPAN-I proton moments to characterize the bulk speed and the temperature tensor. We use the magnetic field measured by FIELDS to rotate the proton temperature tensor into magnetic-field aligned coordinates and thereby separate the parallel and perpendicular internal energy/pressure terms. Large-amplitude velocity fluctuations can briefly take a fraction of the VDF out of the field of view, which affects all moments, but most adversely affects the SPAN-I proton density moments. Therefore, we instead use density values computed from fits to the Solar Probe Analyzer-Electron (SPAN-E) measurements \citep{whittlesey_solar_2020, halekas_electrons_2020}. We have established an absolute density calibration by comparing the SPAN-E fit values to the electron density from quasi-thermal noise measurements from FIELDS \citep{moncuquet_first_2020}. For all quantities that go into the proton energy flux terms, we utilize median values taken over twenty-minute intervals, to remove the effects of the fluctuations in these terms. For this work, we neglect the alpha particle contribution to the energy flux, as well as the proton heat flux, though both of these terms could have important consequences. 

We next add the gravitational potential energy flux corresponding to the energy required for a proton to escape from the gravitational potential well to infinity 
\begin{equation}
    F_{gravitational}= - n U \frac{m_p G M_{\sun}}{r}.
\end{equation}
Some previous studies have referenced this potential to the solar surface in order to keep all energy terms positive \citep{le_chat_solar_2012, liu_solar_2021}. We prefer to instead reference the gravitational potential energy term to infinity. This choice of convention makes the corresponding gravitational energy flux term negative. 

Rather than adding all the electron energy flux terms directly to the proton energy flux terms to form a single-fluid energy flux equation as typically done in the hydrodynamic approach, we instead utilize a single-species approach. We account for the coupling between the protons and electrons by adding the electric potential energy flux term
\begin{equation}
    F_{electric} = nU e \phi_e.
\end{equation}
We utilize the electron distributions measured by SPAN-E to compute the electric potential, by finding the location of the sunward cutoff in the suprathermal portion of the electron velocity distributions \citep{halekas_sunward_2021, bercic_ambipolar_2021}. This cutoff marks the boundary in phase space between regions populated by trapped electrons and regions that remain unpopulated, since any electrons that could populate these regions have instead escaped outward.

Absent significant collisions or resistive terms, the parallel electric field provides all the coupling between protons and electrons. This electric field corresponds to the ambipolar term in a generalized Ohm’s law. For a constant velocity adiabatic radial expansion with $T_e = T_0 r^{-4/3}$, the ambipolar electric field is $e\mathbf{E}= -\nabla P_e/n= 10/3 \: k_B T_0 r^{-7/3} \hat{r}$, and the electric potential with respect to infinity is then $e \phi_e = 5/2 \: k_B T_e$, which exactly replicates the electron enthalpy term in the electron energy flux. However, for the observed electron temperature dependence, $T_e = T_0 r^{-\alpha}$, with $\alpha \simeq 0.5-0.66$ \citep{halekas_radial_2022, liu_total_2023}, we would instead expect an electric potential of $e \phi_e = (2+\alpha)/\alpha \: k_B T_e$, consistent with that in fact observed by PSP \citep{halekas_sunward_2021, halekas_radial_2022}. One can regard this as a polytropic behavior of the electrons, with $\gamma = 1 +\alpha/2 \simeq 1.25-1.33$, consistent with the electron polytropic index derived from other considerations \citep{dakeyo_statistical_2022}. This well-known non-adiabatic electron behavior presumably results in part from the contribution of other terms, such as the electron heat flux, to the electron energy equation, as well as the departures from an isotropic Maxwellian electron distribution \citep{scudder_theory_1979}. Regardless, the electric potential naturally accounts for all these effects, and communicates the result to the solar wind protons. We therefore do not explicitly add the electron heat flux; however, we do separately keep track of the heat flux (estimated using the same techniques as \citet{halekas_electron_2020}) for comparison purposes. 

The terms considered so far are essentially a single-species energy flux equation for the protons (see, e.g., \citet{lemaire_kinetic_1973}). To this, we add the Alfv\'{e}n wave energy flux carried in outward-propagating fluctuations. As \citet{hollweg_alfvenic_1974} has shown, the wave pressure should primarily affect the ions in the solar wind, so it makes sense to add the wave term to the proton energy flux. We express the wave energy flux in the same way as \citet{chandran_incorporating_2011}, as
\begin{equation}
    F_{wave}=\left( \frac{3 U}{2} + v_A \right) E_w.
\end{equation}
Here the wave energy is
\begin{equation}
    E_w = n m_p \frac{<|\mathbf{z}^{\pm}|^2>}{4}, 
\end{equation}
with the Elsasser variables $\mathbf{z}^{\pm} = \delta \mathbf{v} \mp \delta \mathbf{b}$ \citep{elsasser_hydromagnetic_1950}, and the $\pm$ chosen to capture outward-propagating fluctuations (i.e. the minus sign for radially outward magnetic fields, and the plus sign for radially inward magnetic fields). We calculate the fluctuation terms $\delta \mathbf{v}$ and $\delta \mathbf{b}$ by taking root-mean-squared (RMS) values of vector velocity fluctuations (with respect to the mean velocity) measured by SPAN-I and vector magnetic field fluctuations (with respect to the mean field) measured by FIELDS. We convert the magnetic field term $\delta \mathbf{b}$ to velocity units by dividing by $\sqrt{\mu_0 m_p n_p}$, with $n_p$ estimated from the SPAN-E density. To compute the total mean square of the amplitude of the outward-propagating fluctuations $<|\mathbf{z}^{\pm}|^2>$, we utilize the mean squared $\delta \mathbf{v}^2$ and $\delta \mathbf{b}^2$ values together with the mean cross-helicity $\sigma_c = 2 \delta \mathbf{v} \cdot \delta \mathbf{b}/(\delta \mathbf{v}^2 +\delta \mathbf{b}^2)$. For all of the wave energy calculations, we utilize averages over twenty-minute intervals. This duration is long enough to capture the pervasive large amplitude Alfv\'{e}n waves (including switchbacks), although it may not capture the full energy in the inertial range. However, using a longer time range would increase the risk of mixing flux tubes of different spatial origin. 

For a steady-state expansion of the wind, the total energy flux should have zero divergence, assuming that we have captured all relevant terms. If adjacent flux tubes do not strongly interact with each other, then the parallel energy flux multiplied by the flux tube area should remain constant at all distances. In other words 
\begin{equation}
    (F_{proton}+F_{gravitational}+F_{electric}+F_{wave})*Area = const.
\end{equation}
The same should hold for the proton number flux $n U$. Note however that the action of shear-driven instabilities \citep{ruffolo_shear-driven_2020} could invalidate these assumptions.

\begin{figure}
\epsscale{0.75}
\plotone{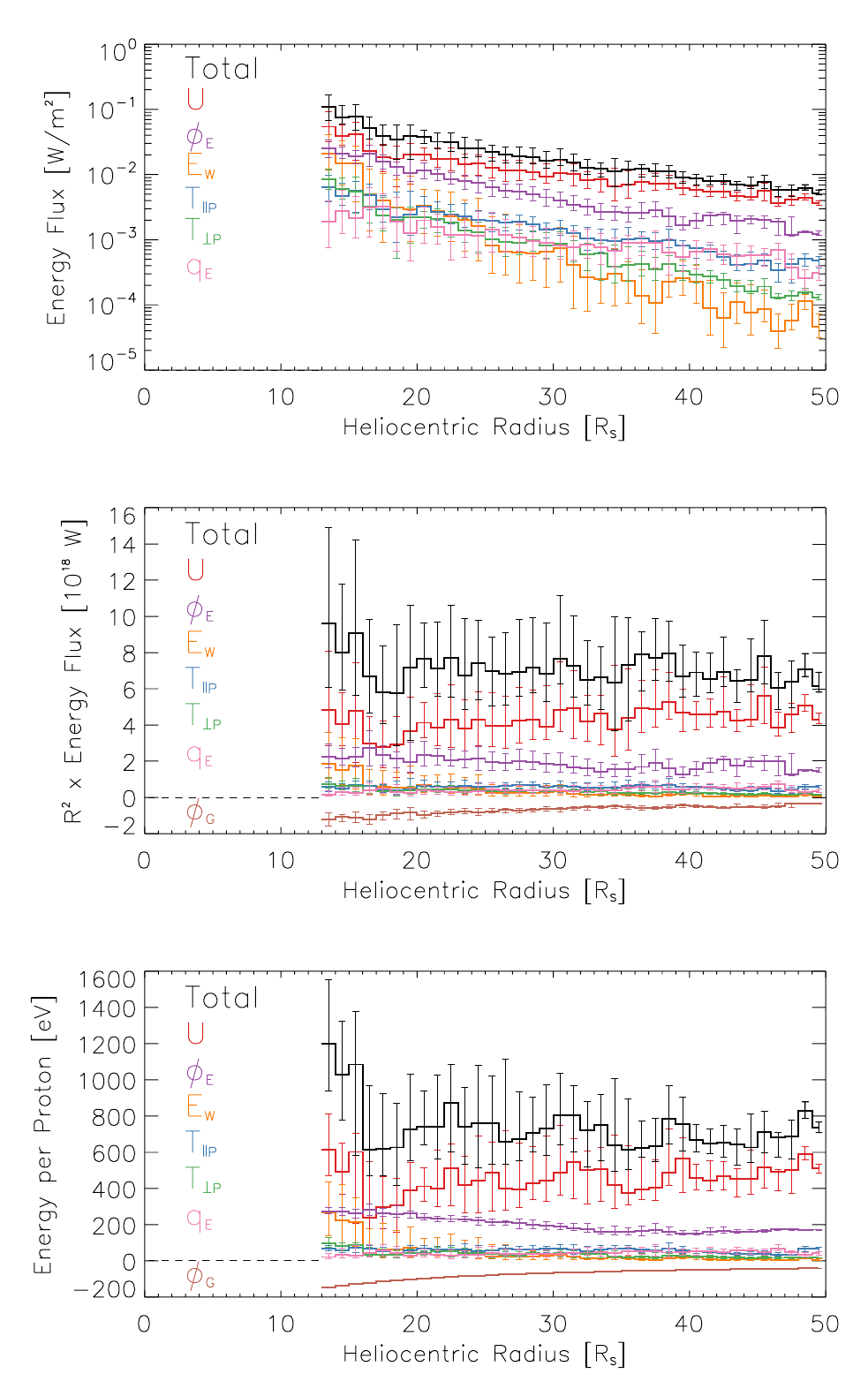}
\caption{Energy flux terms measured by the Parker Solar Probe (PSP) between 13.3 and 50 $R_{\sun}$ on orbits 4-13. In the top panel, colored lines and error bars show the medians and quartiles of the energy flux carried in the form of kinetic energy ($U$, red), electric potential energy with respect to infinity ($\phi_e$, purple), Alfv\'{e}n wave energy ($E_w$, orange), proton parallel and perpendicular internal energy/pressure ($T_{\parallel P}$ and $T_{\perp P}$, blue and green), the total of these terms plus the negative gravitational potential energy with respect to infinity (Total, black), and the electron heat flux ($q_E$, pink), all as a function of heliocentric radius. The second panel shows the same energy flux terms, as well as the gravitational potential energy term ($\phi_G$, brown), each multiplied by the square of the radius. The third panel shows the energy flux divided by the proton number flux (i.e. the energy per proton), for each term.   \label{fig:eflux}}
\end{figure}

We plot the energy flux terms described above as a function of heliocentric radius in Fig. \ref{fig:eflux}. All energy flux terms decrease steeply with distance, so we also show the results of two different normalization methods. Assuming steady state outward flow in a purely radial magnetic field, the total energy flux would decrease as $r^{-2}$, so we compute a normalized energy transport rate by multiplying by $r^2$, as in previous work.  We also employ a normalization that does not rely on the assumption of purely radial expansion, dividing the energy flux by the proton number flux to compute a characteristic energy per proton. With either of these normalization conventions, we find a total quantity approximately constant as a function of distance, consistent with previous work \citep{liu_solar_2021}. The few apparent departures from a constant value very likely result from non-uniform sampling of fast and slow wind streams at different radial distances. In particular, the apparent increase in the total energy per proton at the lowest radial distances results from a higher percentage of faster wind observations in these radial bins. 

In comparing the individual energy flux terms, we find that the kinetic energy flux term dominates over the majority of the radial range, as expected given the significant fraction of the wind acceleration that has already occurred inside of 13.3 $R_{\sun}$. On average, the electric potential energy term contributes the second most energy flux, followed in magnitude by the (negative) gravitational potential energy term. The proton internal energy/pressure terms and the wave energy terms remain smaller over most of the radial range, though the wave energy term becomes appreciable at the smallest radii. The electron heat flux proves smaller than all the other terms at the smallest radii, though it does increase slightly in relative magnitude by 50 $R_{\sun}$.

The mixing of fast and slow streams and the uneven sampling of different wind speeds as a function of radius in the observations shown in Fig. \ref{fig:eflux} make it difficult to draw firm conclusions about the importance of different energy flux terms. To overcome this sampling issue, one can define “wind families” based on some statistical criteria, thus grouping streams with common properties, as done by \citet{maksimovic_anticorrelation_2020} and \citet{dakeyo_statistical_2022}. Instead, we choose to use the total energy flux per proton number flux (energy per proton) to define wind families in this work. This is similar to the grouping by “asymptotic speed” in \citet{halekas_radial_2022}, but it accounts for terms other than kinetic and potential energy. Since the kinetic energy flux dominates the overall energy budget, this also approximately orders the data by wind speed, but it does not require the (incorrect) assumption of a constant wind speed. In fact, since we group winds with similar energy fluxes rather than the same bulk speeds, this allows us to track the evolution and acceleration of the flow speed with radius in each wind family (assuming, of course, that the steady-state zero-divergence condition represents a good approximation for both energy flux and number flux along each flux tube). 

\begin{figure}
\plotone{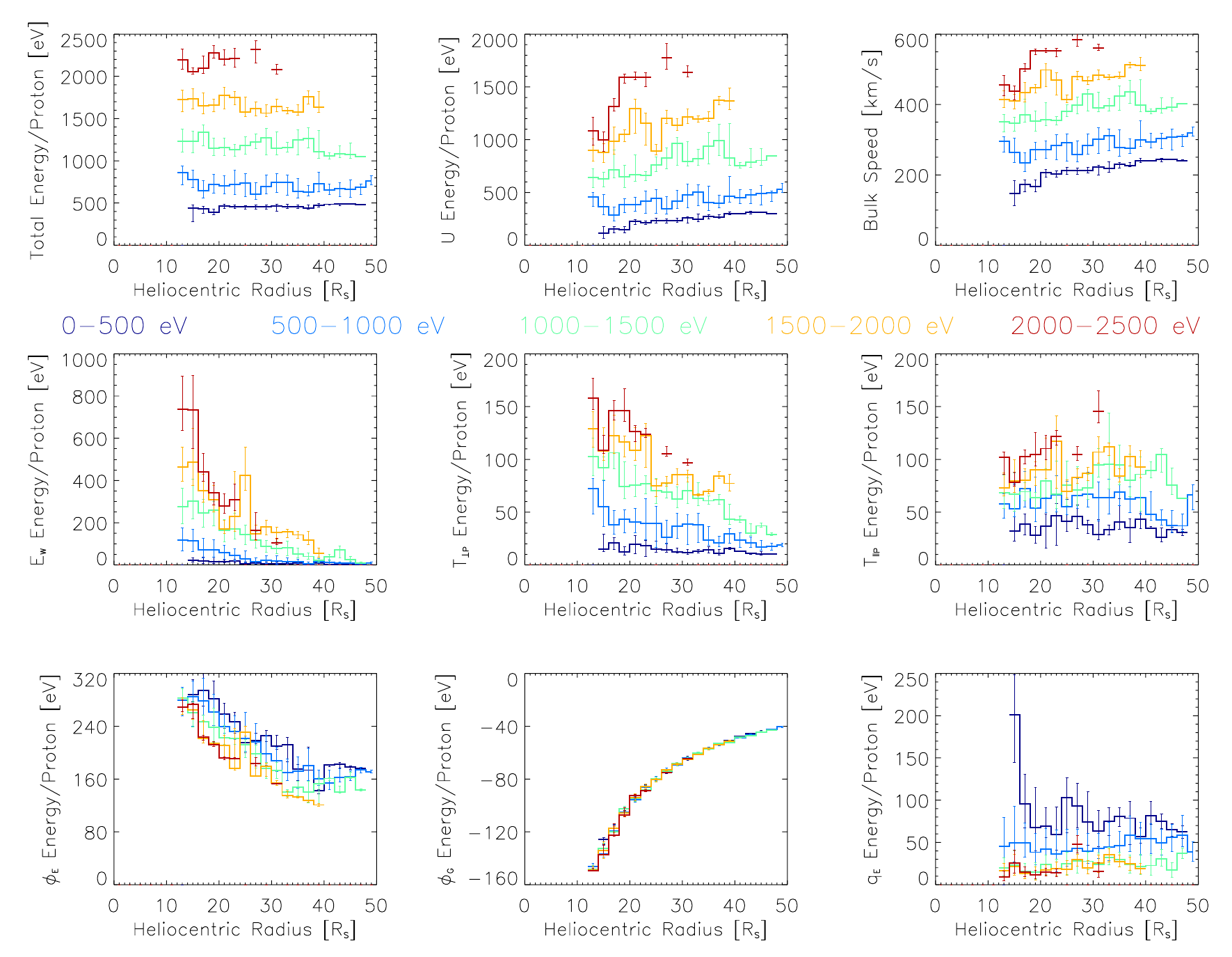}
\caption{Energy flux per proton number flux (energy per proton) for five equal ranges of total energy per proton (0-500, 500-1000, 1000-1500, 1500-2000, and 2000-2500 eV). The nine panels show the medians and quartiles of the kinetic, wave, perpendicular and parallel proton thermal, electric and gravitational potential energy per proton, the total of these terms, and the electron heat flux energy per proton for each of the five ranges, as a function of heliocentric radius. For context, the third panel shows the bulk speed for each range.  \label{fig:eflux_range}}
\end{figure}

We show the results of this grouping in Fig. \ref{fig:eflux_range} for each energy term, and for the bulk velocity for context. As expected, the grouping by total energy per proton clearly orders the kinetic energy per proton and the bulk speed. Within each grouping, we find a characteristic acceleration profile, with significant acceleration outside of 13.3 $R_{\sun}$ in all ranges. Meanwhile, the wave energy and the proton internal energy/pressure terms also correlate with both the total and kinetic energy, and thus with bulk speed. However, the wave and perpendicular energy per proton fall rapidly with radius (especially in the more energetic wind families), while the parallel proton energy varies little with radius. In contrast, the electric potential energy and the electron heat flux energy per proton anti-correlate with total and kinetic energy per particle (and with bulk speed), in agreement with previous PSP observations \citep{halekas_electron_2020, halekas_sunward_2021}. Meanwhile, the electric potential energy per proton falls rapidly with radius, while the heat flux energy per proton varies less. Finally, the gravitational potential energy per proton does not depend on the other energy term, as naturally expected.

\begin{figure}
\plotone{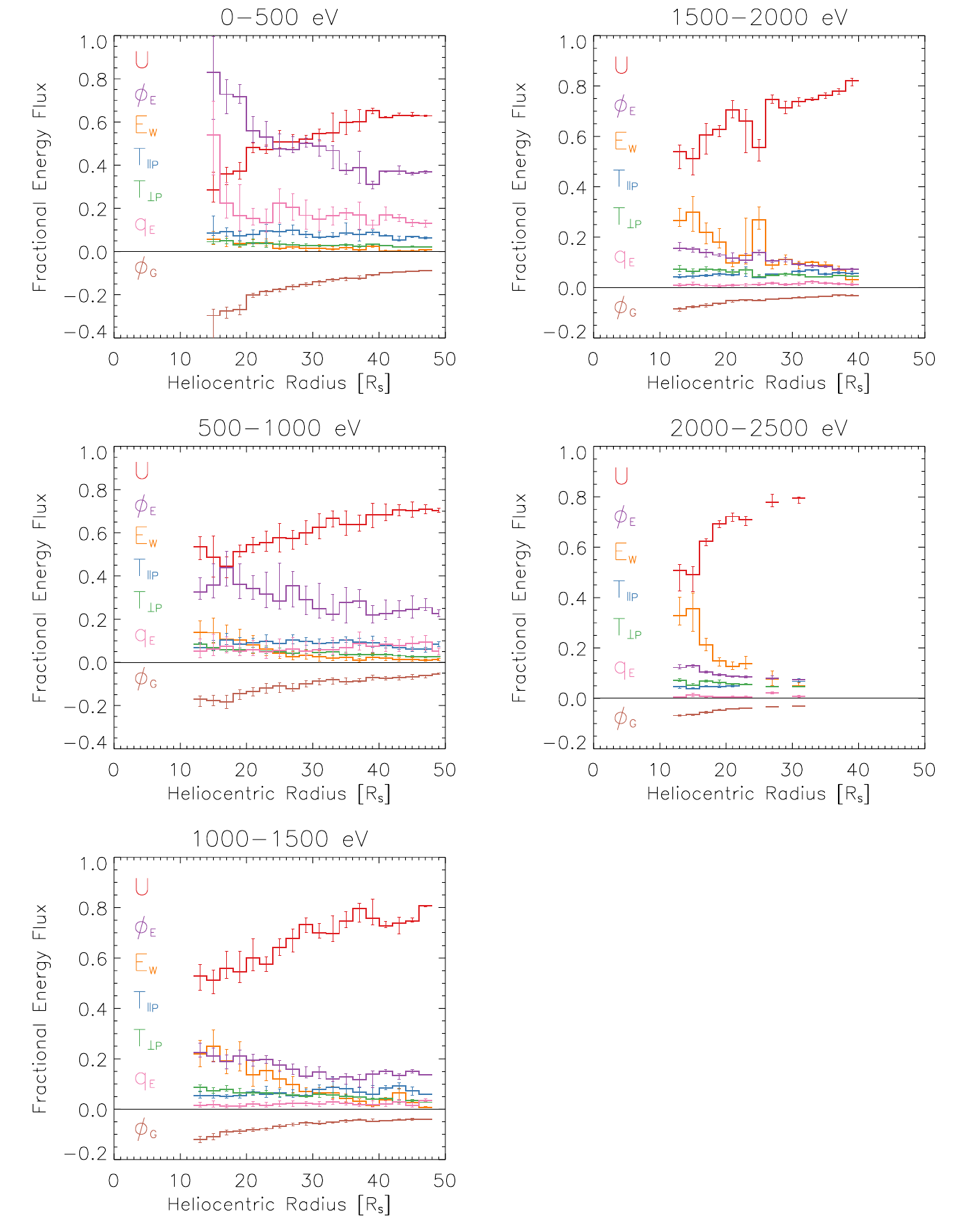}
\caption{Fractional energy fluxes for the same five ranges of total energy per proton as in Fig. \ref{fig:eflux_range}. Each panel shows the median and quartile fractions of the total energy flux carried as kinetic energy, electric and gravitational potential energy, proton parallel and perpendicular internal energy/pressure, and Alfv\'{e}n wave energy, as well as the electron heat flux for comparison, as a function of heliocentric radius.   \label{fig:eflux_frac}}
\end{figure}

The fact that the wave energy, proton perpendicular thermal energy, and electric potential energy per proton decline with radius as the kinetic energy increases makes them natural candidates for wind acceleration energy sources. However, given their different ordering by bulk speed, the relative contribution of these terms will clearly vary between slow and fast winds. To better show the relative importance of the various terms in the different wind families, we display the same data in a different format. In Fig. \ref{fig:eflux_frac}, we show all energy flux terms for each wind family, as a fractional contribution to the total energy flux (presumed constant). This format better reveals the different acceleration profiles and the relative contributions of the various energy sources in the slow and fast winds. 

In the slower wind families, the kinetic and potential energy terms clearly dominate the energy budget. This suggests that the electric potential contains enough energy to provide the main energy source that both overcomes gravity and provides a continuing acceleration to the solar wind protons in the slow wind, in agreement with previous results \citep{horaites_heliospheric_2022, halekas_radial_2022}. However, we emphasize that this energy argument does not definitely show which mechanism or mechanisms actually act to convert energy into acceleration. In the slowest wind family, the electric potential energy even exceeds the kinetic energy at the innermost radius, indicating that the majority of the energy transfer occurs after 13.3 $R_{\sun}$. For the rest of the wind families, the kinetic energy exceeds the other energy terms over the entire radial range considered here, indicating that the majority of the acceleration has taken place already by 13.3 $R_{\sun}$. As one would expect, the slow wind acceleration profiles closely match those analyzed by \citet{halekas_radial_2022}, which previously considered the kinetic and potential energy terms. These observations therefore also remain consistent with the conclusions of \citet{halekas_radial_2022} that an inward extrapolation of the electric potential contains enough energy to account for the entire acceleration of the slow wind. Given its smaller magnitude, the electron heat flux energy term apparently has little appreciable contribution except in the slowest wind, at least in radial range considered here. In any case, we presume that the electric potential term already largely accounts for any coupling of the electron heat flux to the protons. 

In the faster wind families, a rather different picture emerges. The electric potential, while still significant, no longer dominates the energy budget. Instead, the wave energy term takes over as the second largest energy term (behind the kinetic energy). We can therefore identify the wave pressure as the dominant energy source, and thus potentially responsible for the continuing acceleration of the faster wind streams in the radial range considered here, in agreement with a variety of theoretical and simulation work \citep{zank_waves_1992, tu_two-fluid_1997, holst_data-driven_2010, chandran_incorporating_2011, zank_theory_2017, zank_theory_2018, zank_spectral_2020, reville_role_2020}. We note that this observation does not necessarily imply direct conversion from wave energy to kinetic energy. Instead, this could occur through an intermediary step, such as heating communicated to the protons through the proton internal energy/pressure terms and then converted to acceleration. However, this analysis identifies the wave term as the likely ultimate energy source. 

Given the significant contribution of wave pressure apparent at 13.3 $R_{\sun}$, this term very likely also plays a significant role at smaller radii, consistent with the observations of \citet{telloni_first_2023}. On the other hand, additional terms, or even additional processes, may also become more important closer to the Sun. Furthermore, magnetic reconnection may play an important role in producing the initial energy flux in one or more terms. Regardless, significant acceleration must clearly take place before 13.3 $R_{\sun}$ in the faster streams.

\section{Fast Radial Scans} \label{sec:scans}

Though the analysis presented above allows us to draw some important preliminary conclusions, it also has some drawbacks. Since we rely on statistical arguments, we run the risk of grouping different wind streams that do not in fact result from the same mix of physical processes. If we could instead track the evolution of a single stream as it flowed out from the Sun, we could come to even stronger conclusions. Unfortunately, with a single spacecraft, we can never accomplish this (though we note that others have started to investigate PSP-Solar Orbiter conjunctions \citep{telloni_evolution_2021, telloni_linking_2022, reville_flux_2022}, which may help achieve this goal). Nonetheless, thanks to the orbital dynamics that the PSP spacecraft experiences near perihelion, we do often have the opportunity to sample the same streams at multiple radii \citep{shi_alfvenic_2021, badman_prediction_2023}. These repeated crossings of a single flux tube have been dubbed “fast radial scans” \citep{fox_solar_2016}. 

Unfortunately, we have not yet developed a reliable method of calculating all of the important energy flux terms outside of $\sim \! 50 R_{\sun}$. The SPAN-I observations no longer capture the full ion distribution at larger radii, and the sunward deficit in the electron VDF that allows us to infer the electrostatic potential becomes much more difficult to identify \citep{halekas_sunward_2021}. Nonetheless, we can at least still characterize the kinetic energy flux at larger radii, and compare this to the total energy flux available at smaller radii. To accomplish this task, we have constructed a merged data set of density and bulk speed extending to larger radii, created by combining the SPAN-I observations near perihelion and the Solar Probe Cup (SPC) observations \citep{case_solar_2020} wherever we do not have SPAN-I coverage and the SPC data has an acceptable quality flag.   

Mapping a stream from one heliocentric radius to the next presents several challenges \citep{badman_magnetic_2020, telloni_evolution_2021, macneil_statistical_2022, badman_prediction_2023}. While one can simply identify times where the spacecraft crosses the same heliographic longitude, this does not correctly account for the finite travel time of the wind. In the time that it takes the wind to travel from a smaller to a larger heliocentric radius, the Sun rotates, and thus we must shift the observations at larger heliocentric radii in longitude to correctly line them up with observations at smaller heliocentric radii. The simplest version of this involves taking the observed speed at the larger radii and ballistically mapping back to the smaller radii. However, this does not account for the ongoing acceleration of the wind. So, we instead use an iterative process to better align the streams. We first ballistically back-map to the inner radii using the observed bulk speed at the outer radii to obtain a first match between the streams, as a function of the effective source longitude at the minimum heliocentric radius during the segment of interest. We then compute the average of the speed at the inner and outer radii (nearly always smaller than the bulk speed at the outer radii), and redo the mapping. We iterate this process three times, which we have empirically found to lead to convergence. We do not attempt to match up the observations at the different radii in terms of latitude, and indeed the PSP trajectory varies by several degrees in heliographic latitude over the perihelion segments, leading to some unavoidable amount of mismatch. 

\begin{figure}
\epsscale{0.9}
\plotone{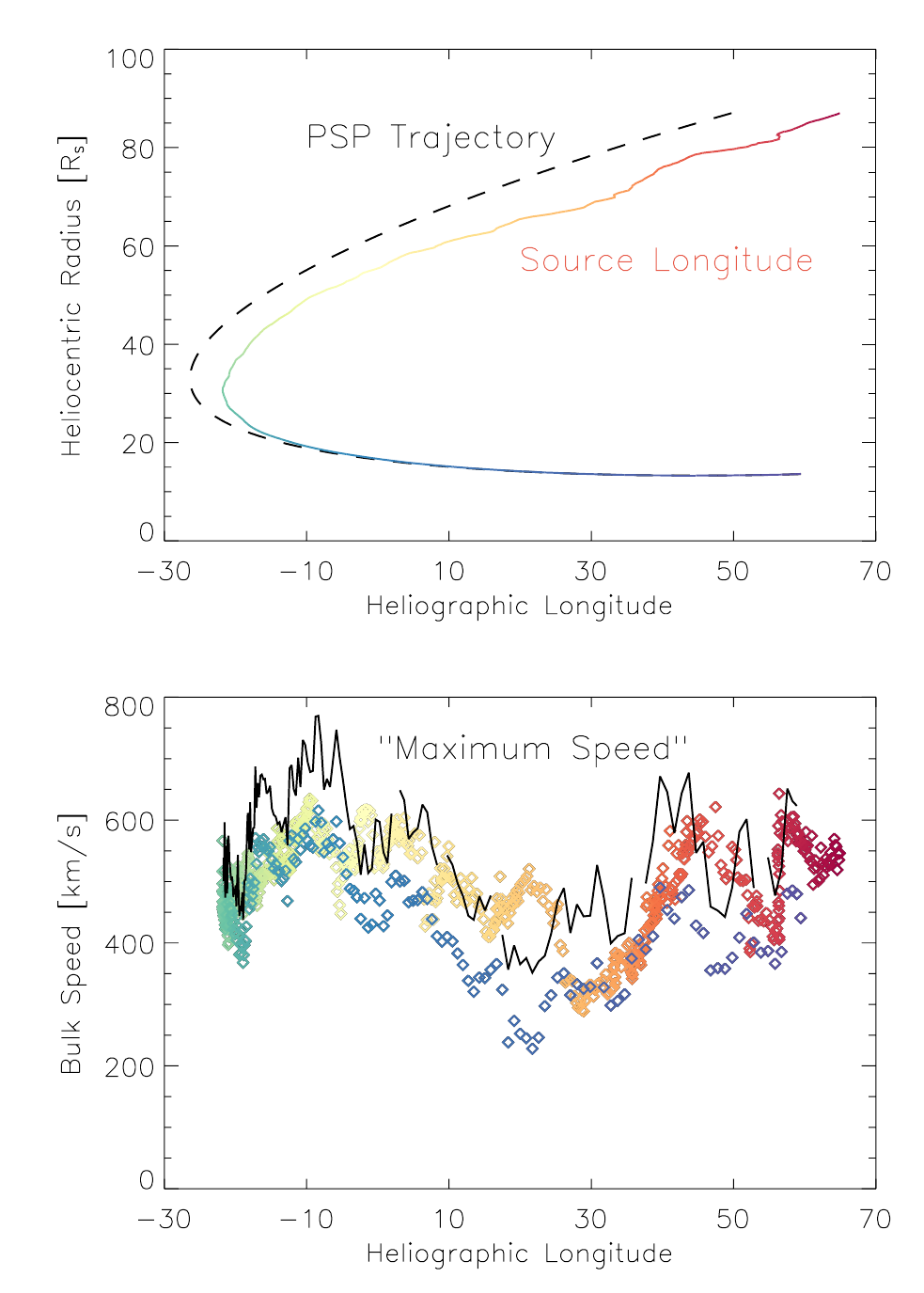}
\caption{Fast radial scan observations from the inbound segment of Encounter 10. The top panel shows the PSP trajectory in terms of heliographic longitude and heliocentric radius, together with the “source longitude” corresponding to the ballistically mapped longitude of the corresponding wind stream at the minimum heliocentric radius, iteratively determined as described in the text. The bottom panel shows the bulk speed measured along this trajectory, with diamonds colored by the same code (corresponding to time) as the locations in the top panel. The solid line shows the “maximum speed”, corresponding to the expected bulk speed obtained by converting all other energy flux terms in the wind to kinetic energy flux. \label{fig:e10_radial}}
\end{figure}

We show an example of this mapping process in Fig. \ref{fig:e10_radial}, for the inbound segment of Encounter 10. We find a reasonably good match between the general characteristics of the observed speed profiles at the inner and outer radii on scales of tens of degrees, suggesting that PSP broadly samples the same streams at the two radii. However, we note only moderate agreement at smaller angular scales, suggesting the presence of some combination of micro-structure in the streams (i.e., due to latitudinal variations), temporal dynamics, and/or inaccuracies in our ballistic mapping process. We also show in Fig. 4 the “maximum speed” that the wind would reach if we could convert all energy flux terms at the lower radii to kinetic energy flux at the outer radii. Given a perfect mapping process and a comprehensive accounting of energy flux terms, one would expect this “maximum speed” to match or exceed the observed speed at all radii. While some short intervals violate this constraint, the majority of the observed data from this orbit segment appear to satisfy this condition, at least to within the accuracy of the mapping process. Therefore, this type of analysis provides a consistency check (albeit an imperfect one) for the more statistical arguments described above.

\begin{figure}
\plotone{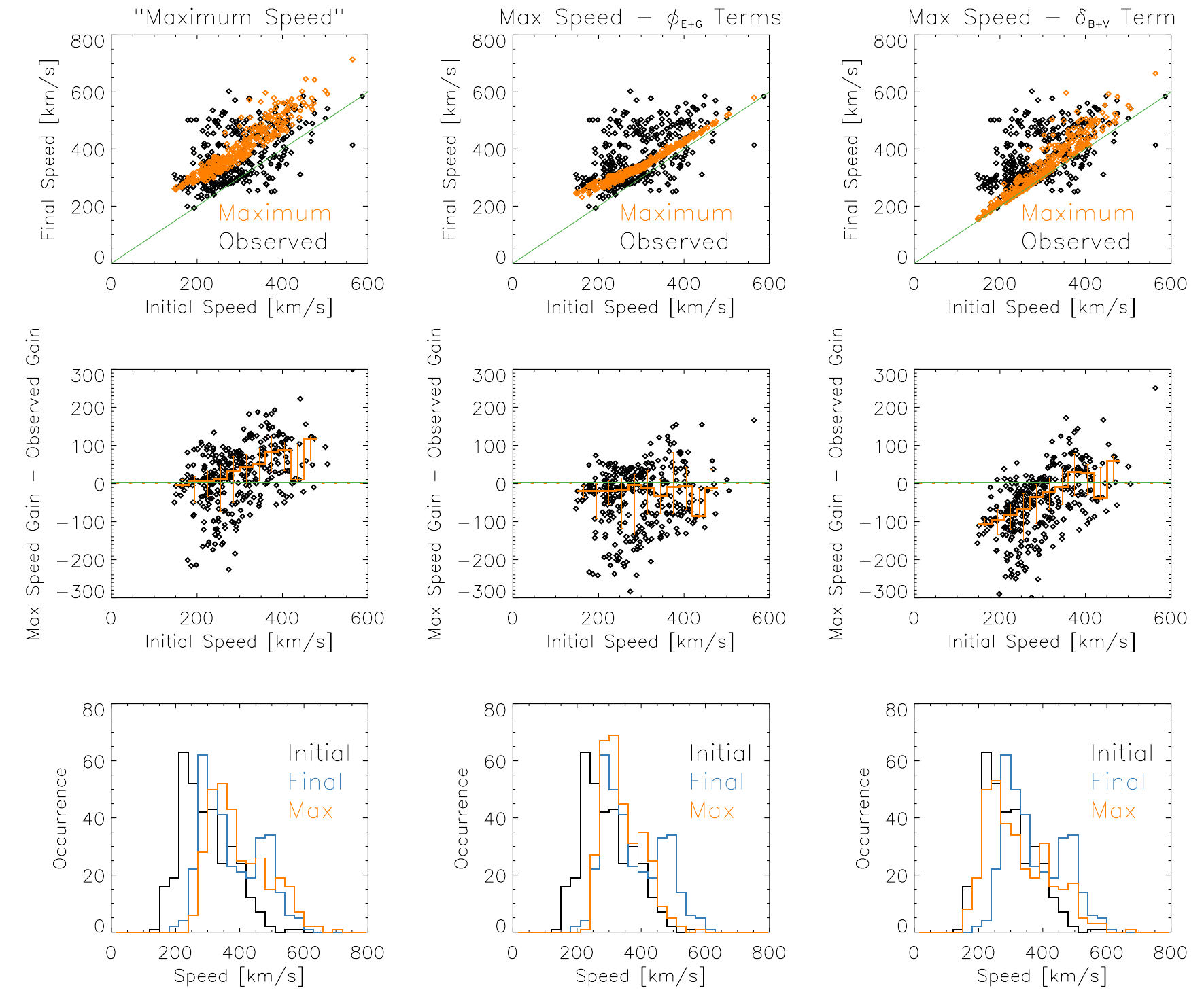}
\caption{Summary of fast radial scan observations from Encounters 4-13. The top three panels show the measured speed at the larger heliocentric radius (“final speed”) vs. the measured speed at the smaller heliocentric radius (“initial speed”), for streams with the same inferred source longitude (one black symbol per degree of longitude). For comparison, orange points show the maximum speed obtained by converting all other energy flux terms to kinetic energy flux, by converting only potential energy flux terms to kinetic energy flux, and by converting only wave energy flux to kinetic energy flux. The middle three panels show the difference between the maximum speed gain and the observed speed gain (i.e. the difference between the orange points and the black points from the top three panels) for the same three cases (black symbols), as well as medians and quartiles thereof (orange lines and error bars). The bottom three panels show the occurrence of the initial (black) and final (blue) bulk speeds, together with the occurrence of the maximum final speed for the same three cases (orange).    \label{fig:all_radial}}
\end{figure}

We repeat the analysis of Fig. \ref{fig:e10_radial} for twenty selected intervals from Encounters 4-13. We choose each interval individually, requiring that the observations at the inner and outer radii have the same magnetic polarity, and that each segment has adequate data quality and coverage. We then collate the results into a single data set. To increase the chances of successful matching of the streams at the inner and outer radii, we require that the ratio of the proton number flux multiplied by the radius squared (expected constant for purely radial steady-state flow) lies between 0.67 and 1.5. We show the resulting refined data set in Fig. \ref{fig:all_radial} 

In agreement with the Encounter 10 test case shown in Fig. \ref{fig:e10_radial}, we find that the maximum speed calculated from the bulk speed at the inner radii and the total energy flux exceeds the observed speed at the outer radii in an average sense, and also for the majority of individual observations. The exceptions may result from inaccurate mapping, cases where other energy terms not yet considered play a role, and/or cases where temporal dynamics or stream-stream interactions affect the acceleration profile. 

We further proceed to analyze the contributions from the potential energy terms and from the wave energy separately. In agreement with our statistical results, we find that the electric potential contains enough energy to explain the majority of the observed acceleration for the lower speed streams, while the wave energy contains enough energy to explain the majority of the observed acceleration for the higher speed streams. Thus, the radial scan analysis proves entirely consistent with the results of the statistical analysis described above. 

As a last check, we investigate the distribution of observed speeds at the initial and final radii. We find that the initial distribution of speeds has a single peak, while the final distribution of speeds appears to have some bimodality. If we compare to the contributions of the individual energy terms, we find that the electric potential appears capable of explaining the lower peak of the frequency distribution at larger radii, but not the upper peak; on the other hand, the wave energy appears capable of explaining at least a portion of the upper peak, but not the lower peak. While not in any way conclusive, this observation may indicate that these two disparate energy sources work to contribute to the observed dichotomy between slow and fast solar wind.

\section{Conclusions and Implications} \label{sec:conc}

 The PSP observations allow us to at least approximately quantify the steady-state contribution of the proton pressure, the electric potential, and the wave pressure to the solar wind between 13.3 and $\sim \! 100$ solar radii ($R_{\sun}$). In agreement with previous work, we find that the electric potential contains enough energy to fully explain the acceleration of the solar wind protons in the slower wind streams in this radial range. This conclusion is not necessarily in conflict with wave/turbulence driven models, which also naturally include the plasma pressure gradient, a portion of which is communicated to the solar wind protons by the electric potential. Furthermore, we find that the wave pressure plays an increasingly important role in the faster wind streams, also in agreement with a substantial body of theoretical and simulation work. We find that the combination of the electric potential and the wave energy, together with smaller contributions from other terms, contains enough energy to satisfactorily explain the continuing acceleration of the solar wind protons in both slow and fast wind streams beyond 13.3 $R_{\sun}$.   

Many processes may contribute to the acceleration of the solar wind, and their relative importance likely varies as a function of heliocentric distance. In this work, we have investigated the energy budget for solar wind proton acceleration over a substantial radial range. However, we freely acknowledge that we have not addressed the most important portion of the acceleration, during which the solar wind accelerates from the corona and passes through the sonic critical point. While one could attempt to "downward-continue" our results to the corona, we acknowledge that the relative importance of the various energy terms may change at smaller distances, and other terms and/or processes not yet considered in our analysis may become important. Given the recent observations from PSP that suggest a major role for magnetic reconnection in the initial solar wind acceleration \citep{raouafi_magnetic_2023, bale_interchange_2022}, reconnection may at least influence, if not determine, the initial energy balance. 

We also acknowledge that our "follow the energy" approach does not directly reveal the details of the energy transfer mechanisms. While we observe specific energy terms decreasing as the kinetic energy term increases, that does not necessarily imply direct conversion of energy. Instead, intermediary processes may act as temporary sinks and sources of energy. In the case of the electric potential, the mechanism for energy transfer to the solar wind protons appears definitive; however, the mechanisms that ultimately produce the exact (non-adiabatic) electron pressure and thus the electric potential structure we observe still remain uncertain. In the case of the wave energy, meanwhile, even the details of the mechanism or mechanisms that ultimately convert wave energy to flow kinetic energy still remain uncertain. 

Despite the limitations of the current analysis, it does provide definite constraints for theoretical frameworks and simulations. Any slow-wind model should involve an important role for the energy contained in the electron pressure gradient and the associated ambipolar electric field, at least in the radial range considered in this study. Meanwhile, any fast-wind model should involve a major role for Alfv\'{e}nic fluctuations and the associated pressure, at least in the radial range considered in this study. While other processes may (and almost certainly do) play a significant role, the PSP observations indicate that we must account for these two terms.    

\begin{acknowledgments}
We acknowledge support from the Parker Solar Probe (PSP) mission and the Solar Wind Electrons, Alphas, and Protons (SWEAP) team through contract NNN06AA01C. PSP was designed, built, and is now operated by the Johns Hopkins Applied Physics Laboratory as part of NASA’s Living with a Star (LWS) program. JSH also acknowledges additional support from the LWS program through NASA grant 80NSSC22K1014. JLV acknowledges support from NASA PSP-GI grant 80NSSC23K0208. An early version of this work benefited greatly from thoughtful discussions within the ``Heliospheric Energy Budget: From Kinetic Scales to Global Solar Wind Dynamics'' team led by M.~E. Innocenti and A.~Tenerani at the International Space Science Institute (ISSI).
\end{acknowledgments}

\bibliography{references}

\end{document}